# AI/ML-Driven Surface Plasmon Resonance (SPR) and Spectroscopy: Materials Interfaces and Autonomous Experiments

Rigoberto Advincula [1, 2]* and Jihua Chen[1]

[1]Center for Nanophase Materials Sciences, Oak Ridge National Laboratory (ORNL)
1 Bethel Valley Road, Oak Ridge, TN 37830

[2]Department of Chemical and Biomolecular Engineering,
University of Tennessee at Knoxville
1512 Middle Dr, Knoxville, TN 37996
* To whom correspondence should be addressed: radvincu@utk.edu

This manuscript has been authored by UT-BatteIIe, LLC, under Contract No. DEAC05-00OR22725 with the U.S. Department of Energy. The United States Government and the publisher, by accepting the article for publication, acknowledge that the United States Government retains a nonexclusive, paid-up, irrevocable, world-wide license to publish or reproduce the published form of this manuscript, or allow others to do so, for United States Government purposes. DOE will provide public access to these results of federally sponsored research in accordance with the DOE Public Access Plan
(http://energy.gov/downloads/doe-public-access-plan).

## Abstract

This review explores the evolution of Surface Plasmon Resonance (SPR) spectroscopy and sensing, transitioning from fundamental studies of adsorption-desorption kinetics to the sophisticated sensing with Electropolymerized Molecularly Imprinted Polymers (E-MIPs). A significant portion of our previous research focuses on the optical properties, electrochromism of polymer dielectrics, and structure-order correlation in polymer brushes and hierarchical ultrathin films. We then address the transformative impact of Artificial Intelligence (AI) and Machine Learning (ML) in data interpretation, culminating in the conceptualization of Self-Driving Labs (SDLs). The importance of generating high-quality training data through high-throughput experimentation (THE) with the SPR is a possibility. These autonomous systems represent the future of materials science, enabling the rapid, closed-loop discovery and optimization of next-generation SPR sensors and analytical methods. This overview highlights the trajectory for integrating conventional experimental design with AI-driven sensing and analytical chemistry across materials and biomedical applications.

## 1. Introduction

The application of artificial intelligence and machine learning (AI/ML) in optics and materials is paramount to future advances in the field.[1–5] By applying AI/ML to surface plasmonics and sensing, it will further optimize sensor design and accelerate analytical tools and real-time and in-situ sensing.[6] It could enhance data analysis and mining. This will lead to faster and more accurate detection (biosensing, imaging). Surface Plasmon Resonance (SPR) and Localized Surface Plasmon Resonance (LSPR) spectroscopy and instrumentation are the methods of choice.[5,7] It can lead to more democratized and powerful diagnostic tools. ML can predict optical responses using well-established optical formalisms, improve predictions of nanostructures for specific functions, and enhance experimental design. It can accelerate complex dielectric and optical simulations, thereby improving signal-to-noise (S/N) performance and cross-sensitivity in such models. By inverse design and a better noise filter, it can improve the figures of merit, better fitting a hypothesis-driven problem. This will open new and novel applications in materials, biological, and biomedical research. Recent reviews of SPR for sensing material advancements, novel data collection, ML, and mathematical tools have been conducted.[8] Figure 1 showcases the applications of some important AI/ML algorithms in SPR sensing.[6]

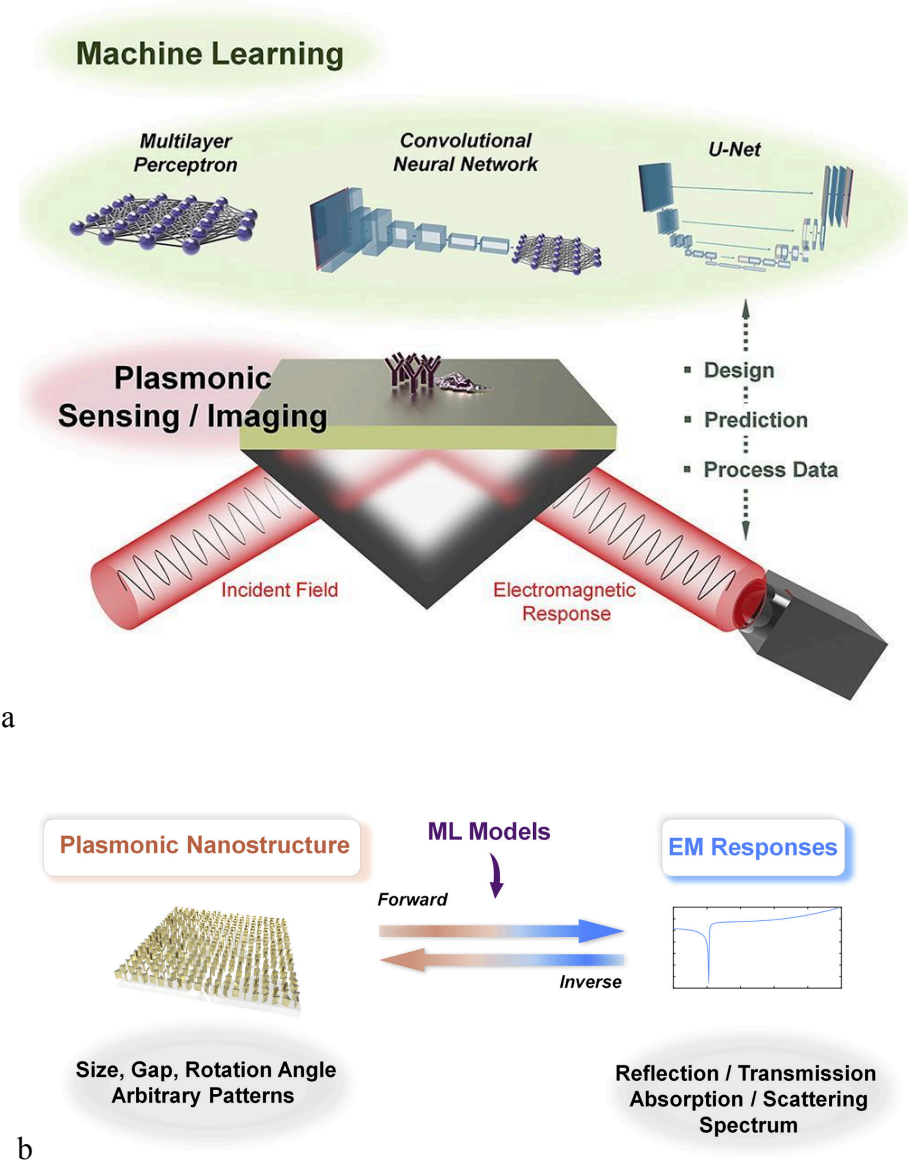

Figure 1. a. Schematic of ML for plasmonics in sensing and imaging applications. Plasmonics-based sensor and imaging systems provide robust platforms for biomolecular sensing and imaging with high sensitivity. ML can be used to analyze diverse types of data from systems. It can also be used to predict EM responses of plasmonic structures and to design desired optical characteristics.  b. Schematics of ML implementation for the design of plasmonic nanostructures. ML can be used to estimate EM responses of plasmonic nanostructures (forward) and to design a plasmonic structure with desired EM responses (inverse). The design structure parameters are not limited to specific values, such as size, gap, and rotation angle; arbitrary patterns are possible.(Figure from reference,[6] with slight rephrasing, under a Creative Commons 4.0 license[9])

## 2. SPR

Surface plasmon (SP) or surface plasmon polaritons (SPP) refer to electron density waves formed at the interface between a metal, air, and/or a dielectric material.[5] SPs can be spatially localized with a confined volume and can produce enhanced electromagnetic (EM) fields due to their higher momentum and local field intensities. It obeys the dispersion relation:[8,10–12]

$$k(\omega) = \frac{\omega}{c}\sqrt{\frac{\varepsilon_1 \varepsilon_2 \mu_1 \mu_2}{\varepsilon_1 \mu_1 + \varepsilon_2 \mu_2}}$$

Eqn 1

where k(ω) represents the wave vector, ε the relative permittivity, μ the relative permeability of the glass block (1) and the metal film (2), ω the angular frequency, and c the speed of light in vacuum.

Resonance shifts of SPs depend on ambient conditions and can be used to measure chemical or biochemical events, which are essentially changes in dielectric properties.[10] A common method for harnessing surface plasmon resonance (SPR) is through evanescent waves generated in an attenuated total reflection (ATR) and prism-coupling setup (Kretschmann configuration), also known as SPR spectroscopy (Figure 2).[10] Many instrument configurations can be derived from an open ATR setup using goniometry or a grating coupler.[11] The layer-dielectric model or formalism based on Maxwell's equations, using methods like Finite-Difference Time-Domain (FDTD) or Transfer Matrix Method (TMM), is well-grounded,[13] and the modeling algorithms are well-established.[14] This enables other hyphenated methods, such as emission, scattering, and microscopy, to be coupled with SPR. Different variations of SPR are surface-enhanced Raman spectroscopy (SERS) and surface-enhanced infrared absorption spectroscopy (SEIRA).[15] Also, localization of SP using plasmonic nanostructures, such as nanoholes, disks, and random arrays, produces hot spots that can modulate optical characteristics for applications in imaging, especially with structured illumination microscopy.[16] SPR microscopy (SPRM) in the near fields can exploit resonance conditions, but only with an improved optical setup and detectors.[17–19]

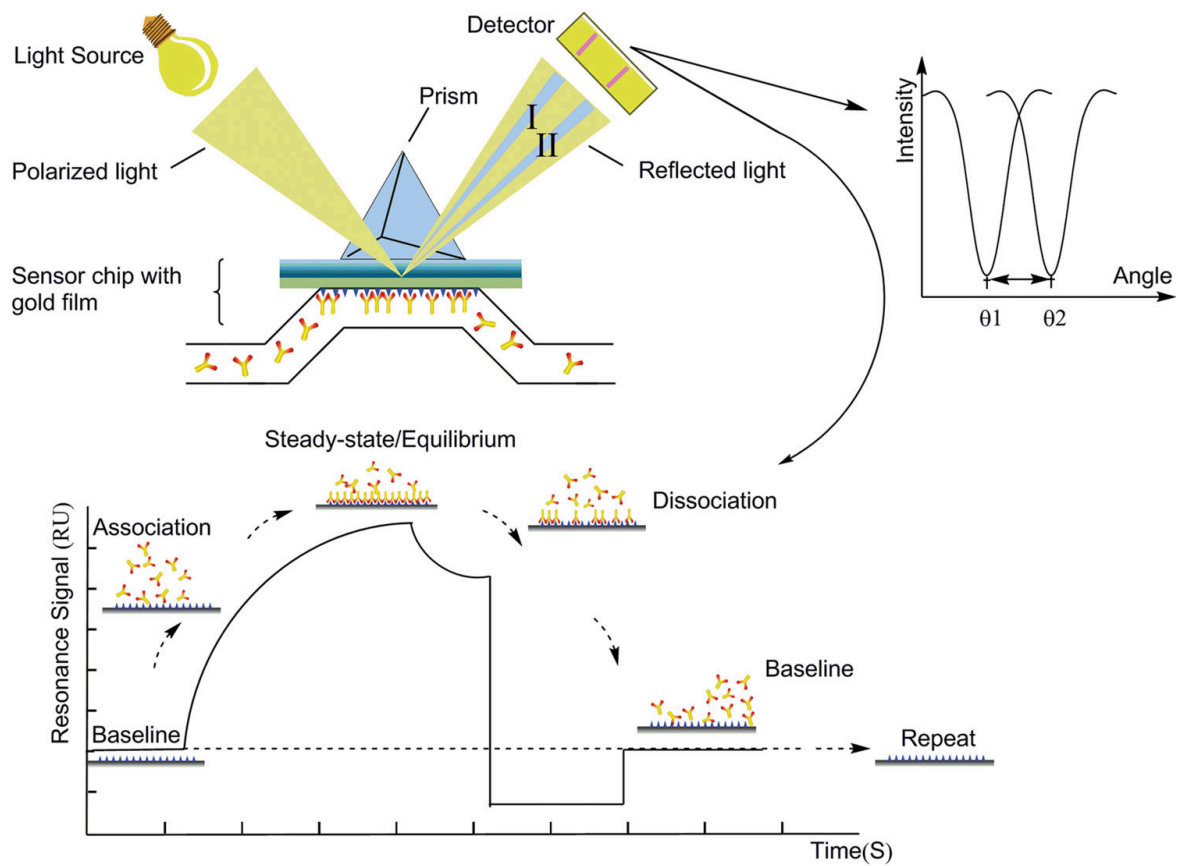

Figure 2. The schematic illustration of surface plasmon resonance (SPR) system. SPR detects changes in the refractive index in the immediate vicinity of the surface layer of a sensor chip. The sensor surface is gold with antibodies attached to it. During the measurement, the chip is irradiated from the bottom with a beam of a wide angle range within that of total internal reflection. The SPR angle shifts (from I to II in the diagram) when biomolecules binding events cause changes in the refractive index at the surface layer. The detector will determine the angle of the intensity decrease. This change in resonant angle can be monitored non-invasively in real time as a plot of resonance signal (proportional to mass change) versus time. (Figure and caption from reference,[20] without changes, under a Creative Commons license.[9])

Given the well-established fundamentals of SPR in physics and photonics, machine learning (ML) can be used to improve the throughput and applicability of SPR techniques.[6,8] ML can use data-driven methods to create and train models, using algorithms without explicit programming. They can be used to predict the outputs of data points and to simplify the interpretation of data, especially with massive data sets.[21] ML can recognize patterns and trends independently of human intervention and is accessed by Python, R, Java, Julia, and MATLAB.[22] ML can thus be used to analyze data, extract meaning and implications, and design plasmonic structures with desired EM responses faster. With inverse design, the instrument and parameters can be tailored to achieve ultrahigh sensitivity and resolution in the presence of SP. Deep learning (DL) can open new possibilities for plasmonic structures and instrument configurations using hyphenated methods and fields.[23] Neural network architectures and algorithms can be used to go beyond conventional designs for a given experiment and to address new phenomena.

## 3. Advances in SPR analytical methods and sensors

The scientific literature is abundant with examples and reviews of SPR, particularly its uses as a sensor and biosensor (Figure 3).[24–26] It also includes many works on optical design and photonics.[2,3,27] In addition, there is a large body of literature on metamaterials and sensing.[28–30] This article is not meant to be comprehensive but focuses on the possibilities of SPR spectroscopy and hyphenated methods when implemented with AI/ML workflows and high-throughput experimentation. First, we look at examples of how SPR has been applied and hyphenated for dielectric measurements, analytical kinetics measurements, electrochemistry, monitoring polymer brush grafting, sensor and biosensor applications, and device optimization. We have previously reported reviews on SPR, electrochemical SPR (EC-SPR) detection, and sensors. But the examples are categorized below:[31–33]

**Adsorption - Desorption Kinetics and Self-assembly:**

One of the early uses of SPR is *in-situ* Investigation of polymer self-assembly solution adsorption and kinetics.[34] A related method is evanescent waveguide spectroscopy for investigating the photodegradation and photochemical crosslinking of polystyrene (PS) thin films.[35] FT-IR imaging and quartz crystal microbalance (QCM) methods complemented the study.[36]

SPR was used to investigate the layer-by-layer self-assembly of nanostructured sexithiophene bolaform amphiphiles, which are relevant to the high-performance organic field-effect transistors (OFET).[37] The thin film self-assembly of organic-inorganic hybrid, metal coordination polymers was examined by *ex-situ* SPR and coupled with UV-Vis absorption and AFM studies.[38] The non-fouling characteristics of an electro-grafted linear-dendron macromonomer were established based on SPR binding studies, in which the oligoetheylene glycol or OEGylated dendron carbazoles were prepared from different dendron generations (G0, G1, G2), and SPR adsorption kinetic studies suggested that the G2 dendrons displayed the best protein resistance.[39]

## Sensors and E-MIP Sensors

In sensing experiments, the sensor element (interface) can define the sensitivity and selectivity along with the transducer element (optical, electrical, spectroscopic, etc.) and read-out devices.[12,16,25] SPR sensors are known to exhibit high figures of merit that combine these elements.[25,27] SPR sensing was established on the oxidation of β-Nicotinamide Adenine Dinucleotide (NADH), which is a key coenzyme in living cells and serves as a mobile electron carrier for metabolism and energy production.[40] The electroactivity of polyaniline/NADH multilayer films was identified in neutral solutions and then compared with QCM. This system is a potential framework for drug discovery and clinical testing.[40] An important contribution to glucose sensing and diabetes diagnostics was reported with the EC-SPR and waveguide-enhanced sensing via N-alkylaminated polypyrrole and glucose oxidase multilayers, through which the SPR /waveguide combination enabled distinguishing between enzymatic and non-enzymatic effects that include adsorption and real-time dielectric changes.[41] We also demonstrated the use of EC-SPR, SPR, and QCM to detect specific cations in azacalix[3]arene−carbazole conjugated polymer network with exceptional sensitivity and selectivity.[42]

The highly selective and sensitive EC-SPR detection of theophylline compared to other drug analytes was revealed using E-MIP based on polythiophenes.(Figure 4) The XPS, QCM, and AFM studies supported the efficacy of molecular templating, as evidenced by improved sensor figures of merit.[43]

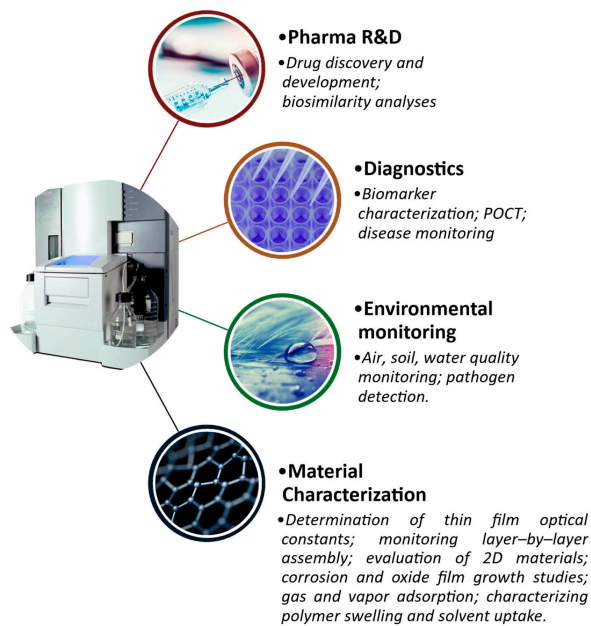

Figure 3. SPR biosensors have found widespread use in pharmaceutical research and development, clinical diagnostics, environmental monitoring, and the characterization of thin films' structural and optical properties. (Figure and caption are from reference [44] without change, under a Creative Commons license.[9]) SELEX refers to Systematic Evolution of Ligands by Exponential Enrichment, an iterative molecular biology selection technique for drug discovery and diagnostics. The goal of SELEX is to find aptamers, or DNA/RNA strands, that can bind to a target molecule. POCT refers to Point-of-Care Testing, or medical diagnostics performed in proximity to patients, often outside a mortar-and-stone laboratory or hospital, using portable devices for fast results to guide timely treatments.

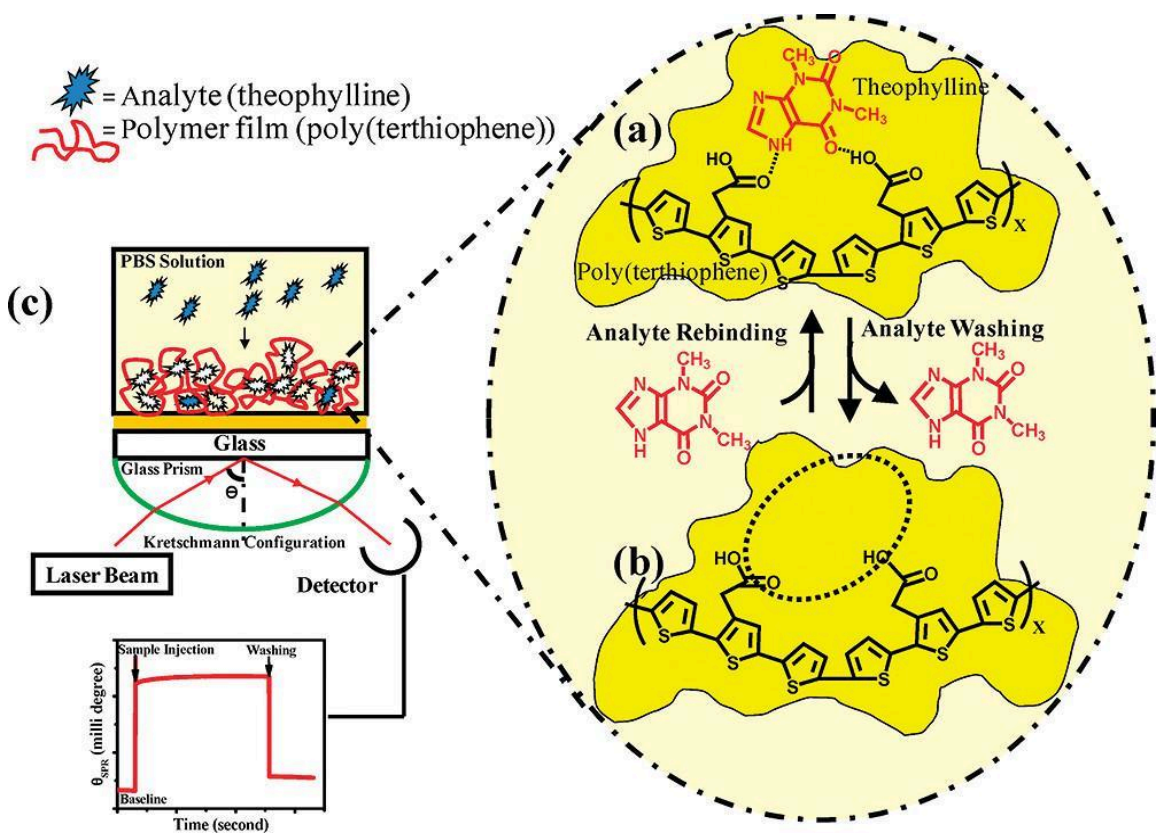

Figure 4. Theophylline detection with SPR and electropolymerized, molecular imprinted polymer (E-MIP). (Figure from reference[43] with author reuse permission from American Chemical Society or ACS. Copyright © 2010 American Chemical Society)

## 4. Our work on SPR

We have reported on the electropolymerization of molecularly imprinted polymers (E-MIPs) for SPR-based sensing of molecules (Figure 5), which involved a pre-polymerization complexation with electroactive monomers such as terthiophene and carbazole (Figure 6).[45] The E-MIP sensing was combined with electrochemical impedance spectroscopy (EIS) to highlight the extraordinarily selective and sensitive detection of Bisphenol A via anodically electropolymerizable carbazole and terthiophene monomer-analyte templating,[46] while SPR identification of dopamine was performed using cathodically electropolymerized MIP of Poly-p-aminostyrene involving vinyl monomers via a radical-anion mechanism (Figure 5b).[47] Even larger polypeptide analytes can be detected by E-MIPs and SPR. The sensing elements for detecting dengue infection were designed with terthiophene monomers and dendrons, resulting in outstanding sensitivity and selectivity for the protein epitope.[48] Our group recently highlighted a broad range of potential applications for plasmonics in biomedicine.[49]

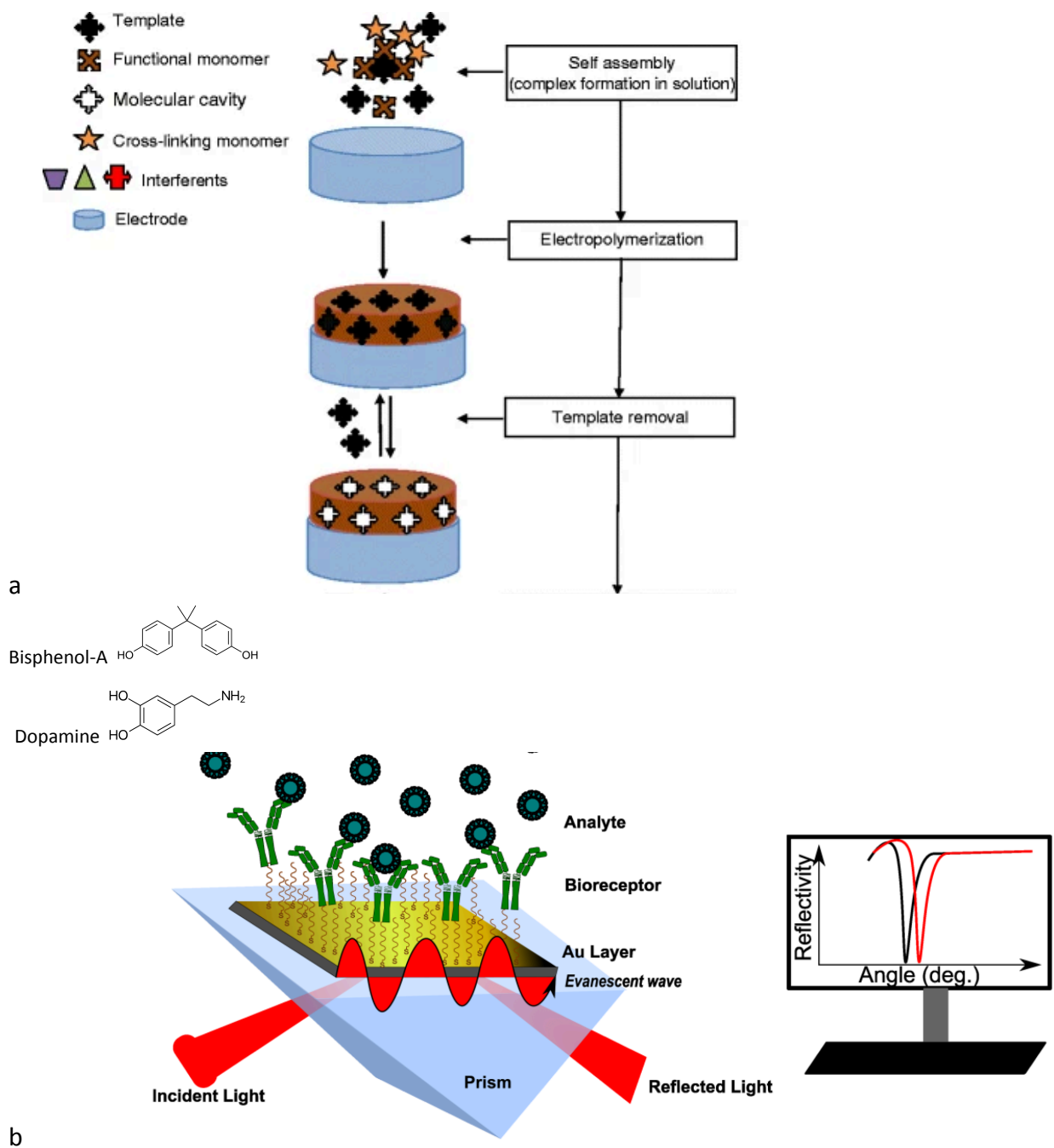

Figure 5. Electropolymerized molecularly imprinted polymers (E-MIPs) for SPR sensing. **a.** General procedure for molecular imprinting with an electroactive functional monomer and typical signal transduction methods employed in detection (Figures and captions from reference[50] with slight rephrases and changes, under open access license.[51]) **b.** A schematic of an SPR sensor with a Kretschmann configuration. Analytes (and bio-analytes) include drug molecules (such as dopamine, shown here), biomacromolecules such as polypeptides and proteins, and other organic small molecules (bisphenol A is shown here as an example). The analyte binding to the immobilized (bio)receptor (i.e., electropolymerized MIPs) causes a change in RI near the surface, which is recorded as a shift in the resonance angle (here, measured at a fixed

wavelength). RI reaches a constant value at equilibrium. (Figures and captions from reference[52] with slight rephrases and change, under a creative commons license.[9])

## Optical Behavior and Electrochromism in Polymer Dielectrics

Figure 7 highlights two widely adopted self-assembly methods for fabricating ultrathin polymer films: the layer-by-layer (LBL) and Langmuir-Blodgett (LB) methods. Incorporating electrochemical surface plasmon spectroscopy (EC-SPR), including surface plasmon field-enhanced light scattering (SPFELS), enabled concurrent optical and electrochemical SPR examinations of self-assembled conducting polymer films.(Figure 8)[53] This setup also allowed optical *in-situ* investigations on the electrochemical polymerization and properties of polyaniline (PANI) films,(Figure 6a) which included their redox behavior and faradaic deposition as observed by cyclic voltammetry (CV).[54] The same setup was utilized to study the work function and tunable hole-injection/transport layers of electrodeposited polycarbazole network in OLED devices.[55]

Poly(3,4-ethylenedioxythiophene) (PEDOT), a well-studied conducting polymer and hole-transporting material (Figure 6b) was investigated using EC-SPR, for which dielectric thin films were investigated for redox and electrochromic behaviors, with the use of two different laser wavelengths highlighting electrochromic behavior and dopant electrolyte contribution.[56] The unique combination of a three-instruments-in-one setup facilitated the simultaneous *in-situ* electrochemical, surface plasmon optical, and atomic force microscopy (EC-SPR-AFM) examination on conjugated polymer electropolymerization. PEDOT was investigated for its time-space resolution when simultaneously using a CV or potentiostatic deposition method.[57]

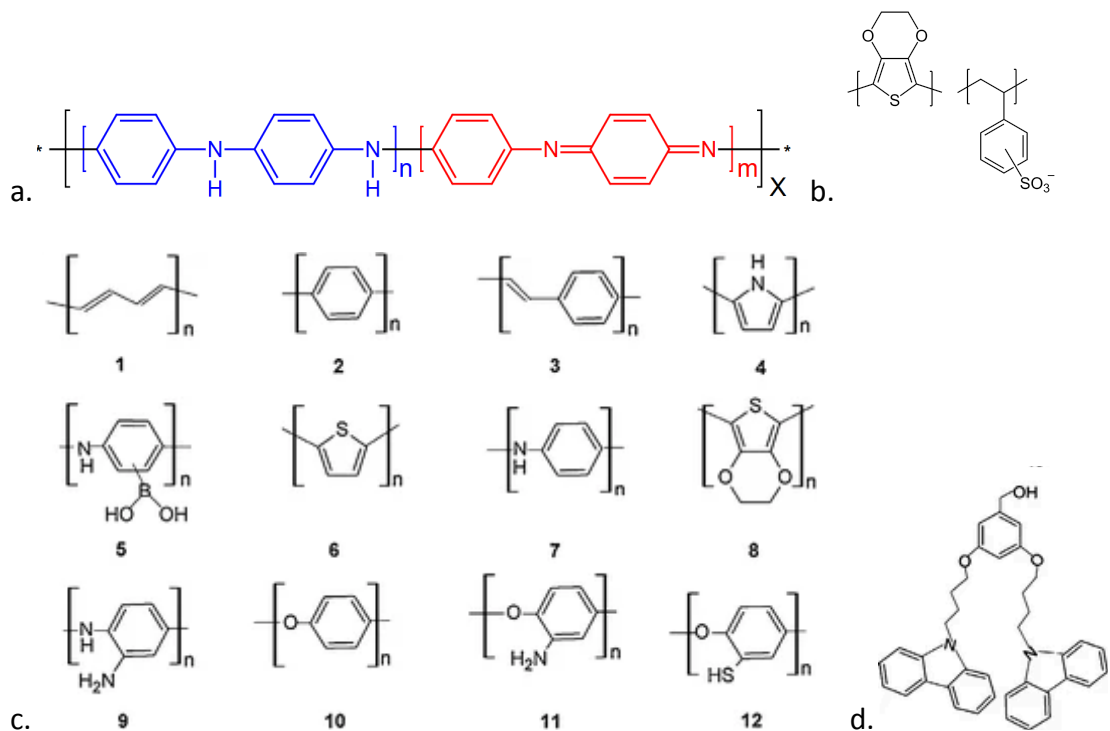

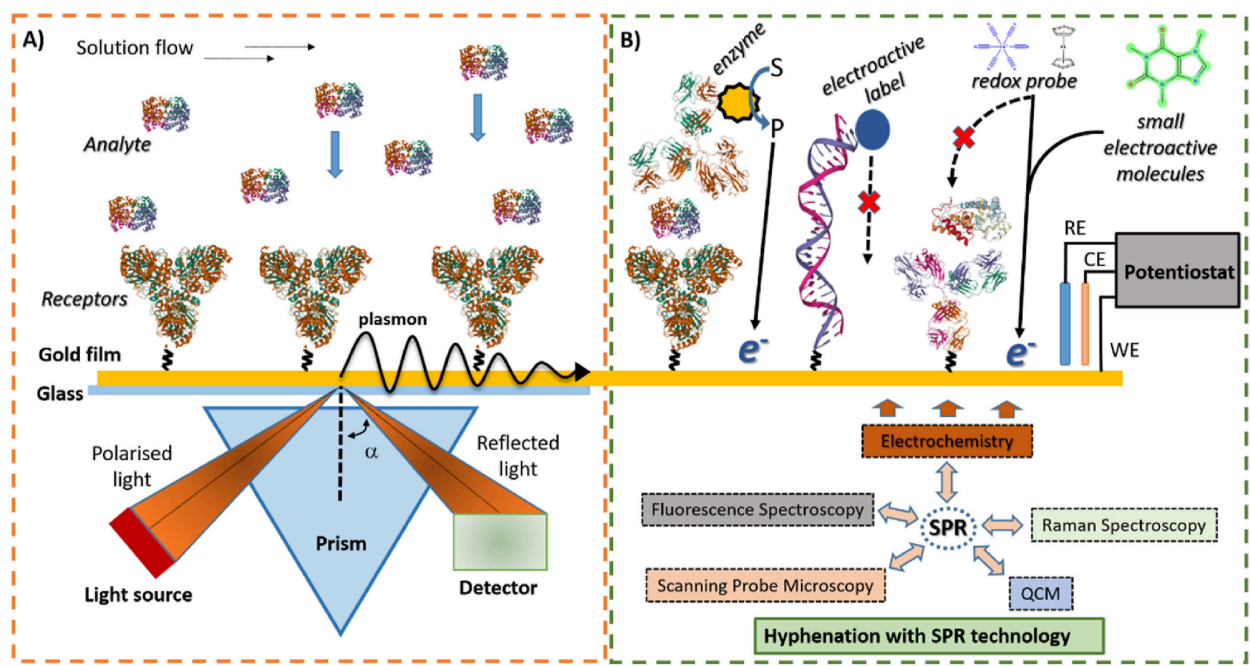

Figure 6. **Top**: Molecular structures of electroactive monomers. **a.** Polyaniline, **b.** PEDOT with polystyrene sulfonate or PSS, **c.** Structural formulas of the most common polymers prepared by electropolymerization. Electronically conducting polymers: polyacetylene 1, polyphenylene 2, polyphenylenevinylene 3, polypyrrole 4, poly(aminophenylboronic acid) 5, polythiophene 6, polyaniline 7, and polyethylenedioxythiophene 8. Electronically nonconducting polymers:
13

polyphenylenediamine 9, polyphenol 10, polyaminophenol 11, and polythiophenol 12, (Figures and captions from reference[50] with changes under a creative commons license.[51]) **d.** The molecular structure of a derivative of carbazole used as a functional monomer for molecular imprinting (Figure and caption from reference[50] with slight changes under a creative commons license.[51])

**Bottom**: A) Schematic illustration of a typical SPR experiment for real-time measuring antibody-antigen molecular interactions, including SPR instrumentation (based on Kretschmann's configuration) and operation mode; B) Examples of SPR combination with other analytical techniques are given, with particular focus on electrochemistry. The setup used for electrochemistry experiments is shown along with common detection schemes used for electrochemical biosensing. (Figures and captions from reference[58] with slight rephrase and change under a creative commons license.[59])

The nanopatterning and assembly of memory devices from layer-by-layer (Figure 7, top) PEDOT and Poly(styrene sulfonate) (PSS) films were augmented with *ex-situ* SPR studies of their deposition.[60] Similar *ex-situ* SPR studies were conducted involving the electrochemical cross-linking and patterning of layer-by-layer, nanostructured polyelectrolyte-carbazole precursor films.[61–63]

Ultrathin films of alternating sexithiophenes and electropolymerizable polycarbazole precursor layers were investigated by EC-SPR, revealing nanostructure and dielectric properties changes during cross-linking for potential hole-transport and charge-carrier mobility applications in solid-state devices.[64] The *in-situ* study on the fabrication and photoisomerization of layer-by-layer films containing azobenzene was carried out by SPR to explore potential applications in liquid crystal display devices.[65] Using SPR, the ultrathin film electrodeposition of polythiophene conjugated polymer network (CPN) through a precursor route was evaluated *ex-situ* and verified by absorption, AFM, and XPS analysis.[66]

The detailed electrochemical and electrochromic behavior of terthiophene and carbazole CPN precursors and their copolymers was investigated by electrochemical-QCM (EC-QCM) and EC-SPR.[67] Ultrathin CPN films of carbazole-functionalized poly(p-Phenylene) were likewise investigated, in which the films were formed by the Langmuir-Blodgett (LB) method (Figure 7, bottom).[68]

The electropolymerization and doping/dedoping characteristics of polyaniline thin films were investigated using a pair of EC-SPR and QCM.[69] The spectroelectrochemical transmittance measurements studied the optical properties of the films as a response of applied potential, thereby determining the electrical, optical, and dielectric features of conjugated ultrathin polymer films *in-situ* as well as in real time.[69]

Signal improvement and SPR tuning in the presence of Au nanoparticles for an Au nanoparticle and polyelectrolyte layer-by-layer film demonstrated pH sensitivity, with swelling and distance-dependent plasmonic band hot spots, while computational considerations based on the Maxwell−Garnett theory were employed to contrast experimental results with the various layer architectures (Figure 9).[70]

An attenuated total reflection and emission investigation with surface plasmon excitation was performed to track the photoisomerization of a layer-by-layer thin film containing azo dyes.[71] Utilizing a technique related to SPR, evanescent waveguiding and photochemical measurement were applied to determine the *cis-trans* isomerization of an all-azobenzene-functionalized dendrimer film.[35]

The emission and multiple surface plasmon (SP) excitations within the prism/Ag/merocyanine LB films were studied in a modified SPR setup, where the emission characteristics were matched with the resonant conditions of SP excitations in a Kretschmann configuration at various thicknesses.[72]

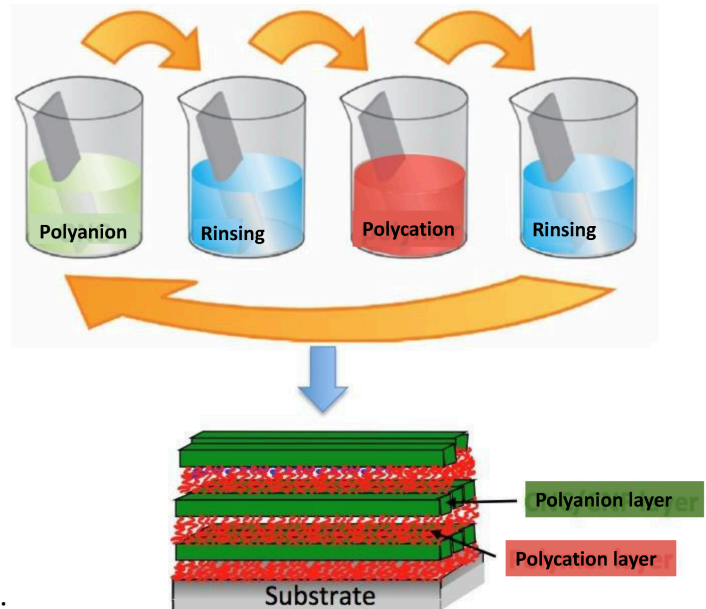

Layer by Layer (LBL):

Langmuir – Blodgett (LB):

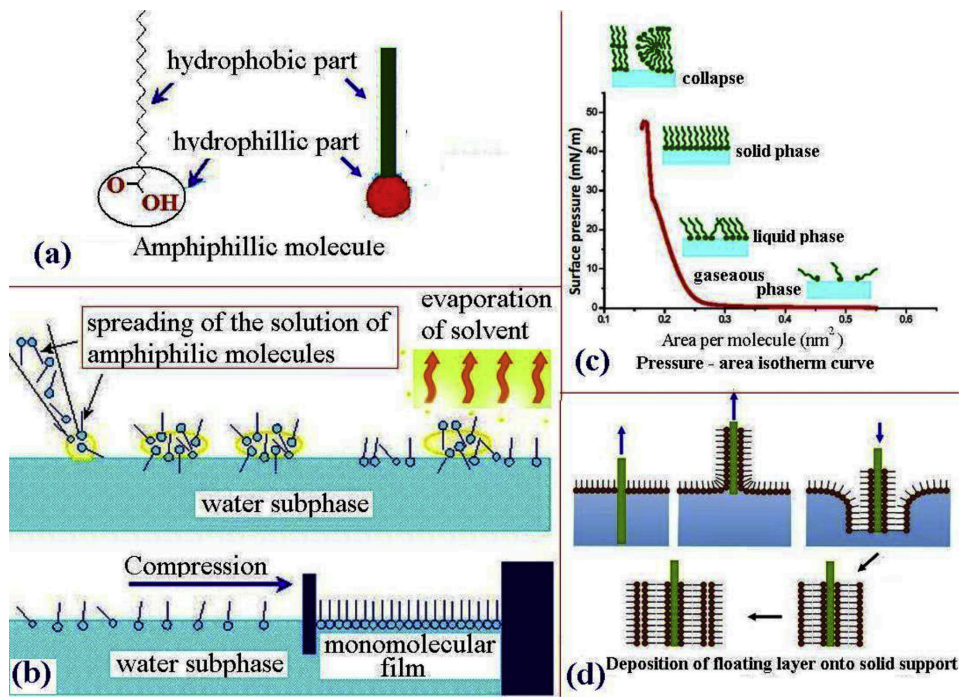

Figure 7. Top: Diagram of the LbL deposition process based on the alternating exposure of a charged substrate to solutions of positively and negatively charged polyelectrolytes. (Figure and caption from reference[73] with no change under a Creative Commons license.[9]) Bottom: Langmuir – Blodgett (LB) technique. (Figure and caption from reference[74] with no change under a Creative Commons license.[9])

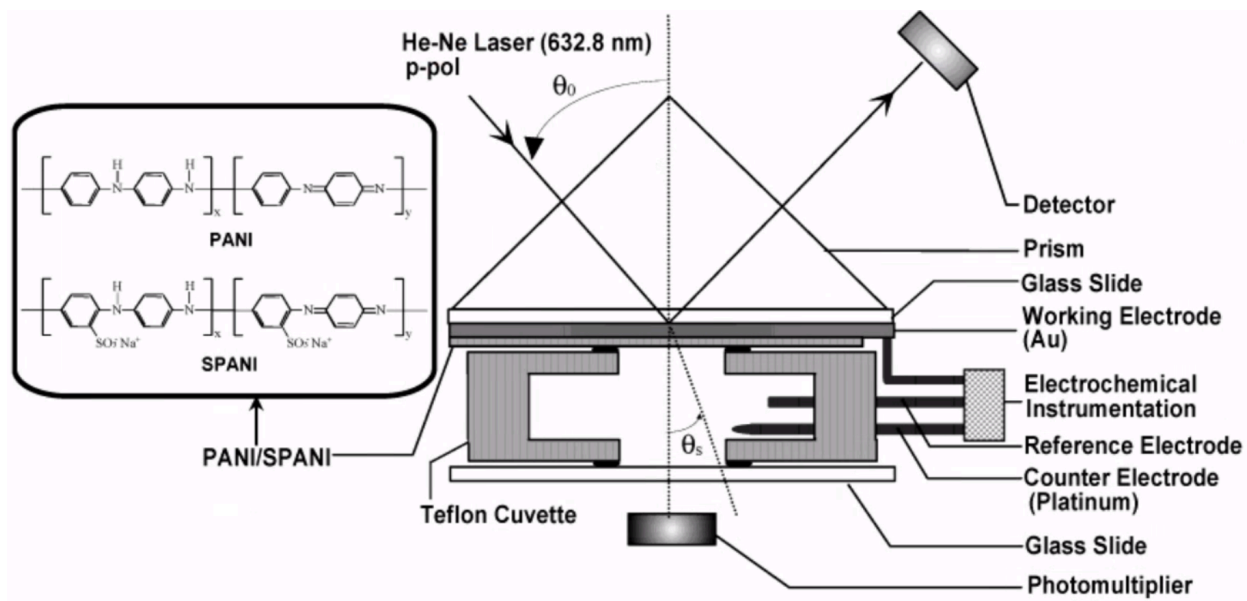

Figure 8. Simultaneously study of Layer-by-Layer assembled conducting polymers with SPR and electrochemical methods (Figure from reference[53] with author reuse permission from ACS. Copyright © 2002 American Chemical Society)

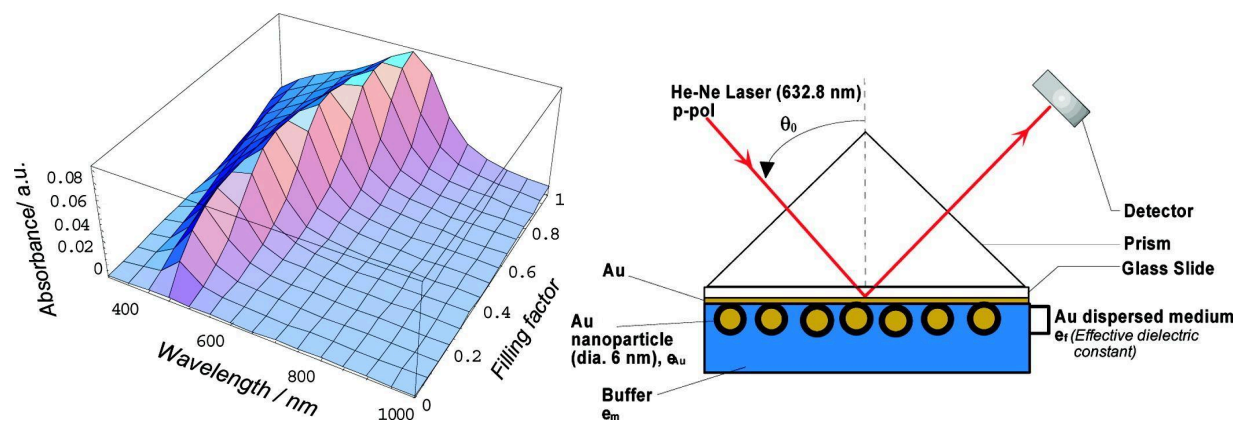

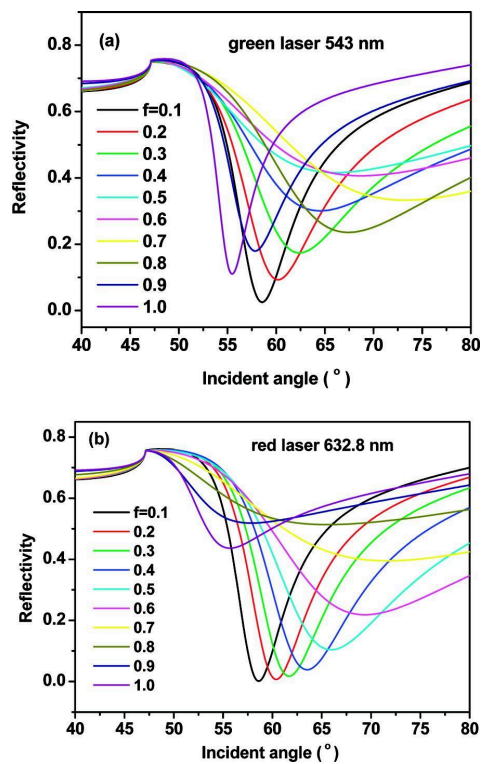

Figure 9. Signal Enhancement and Tuning of Surface Plasmon Resonance in Au Nanoparticle/Polyelectrolyte Ultrathin Films.[70] Top Left: Calculated absorption spectrum of gold nanoparticles dispersed medium as functions of wavelength and filling factor. (A value of 6 nm was used for the diameter of nanoparticles); Top Right: Prism/Au thin film/Au nanoparticle /buffer system used in the calculation; Bottom: Calculated SPR curves of prism/Au thin film/god nanoparticles/buffer system as a function of filling factor of gold nanoparticles under different light sources: (a) green laser, λ = 543 nm; (b) red laser, λ = 632.8 nm. (Figure and caption from reference[70] with slight rephrase under author reuse permission from ACS. Copyright © 2007 American Chemical Society)

**Polymer Brushes and Ultrathin Films**

Table 1 compares polymer brushes and self-assembled monolayers (SAMs). Another important SPR application is for *in-situ* and *ex-situ* characterization of polymer brushes grafted to surfaces (Figure 10). For example, polymer brushes grafted from clay nanoparticles on a planar substrate (via free-radical surface-initiated polymerization, SIP) were characterized using a combination of AFM, X-ray photoelectron spectroscopy (XPS), and *ex-situ* SPR for brushes of different lengths.[75] Homo and block copolymer brushes, poly(styrene) (PS) and poly(ethylene oxide) (PEO), and poly(isoprene)-b-poly(methylmethacrylate) (PI-b-PMMA) were grafted on gold substrates via living anionic surface-initiated polymerization (LASIP) and investigated by SPR, XPS, AFM, and ATR-FTIR to reveal control of polymer brush length and copolymerization.[76]

Vapor deposition was used to evaluate polymerization of an N-carboxy anhydride benzyl glutamate monomer, for which films up to 30 nm were investigated with *in-situ* SPR, AFM, and FT-IR measurements to confirm secondary polymer structures and morphology.[77] More recently, SPR studies were implemented to investigate polymer brush formation by surface-initiated photo-electron transfer RAFT (SI-PET-RAFT) through electrodeposited macroinitiator ultrathin films as a route to control polymer brush length and grafting density for potential applications in biomedical devices, including sensing.[78]

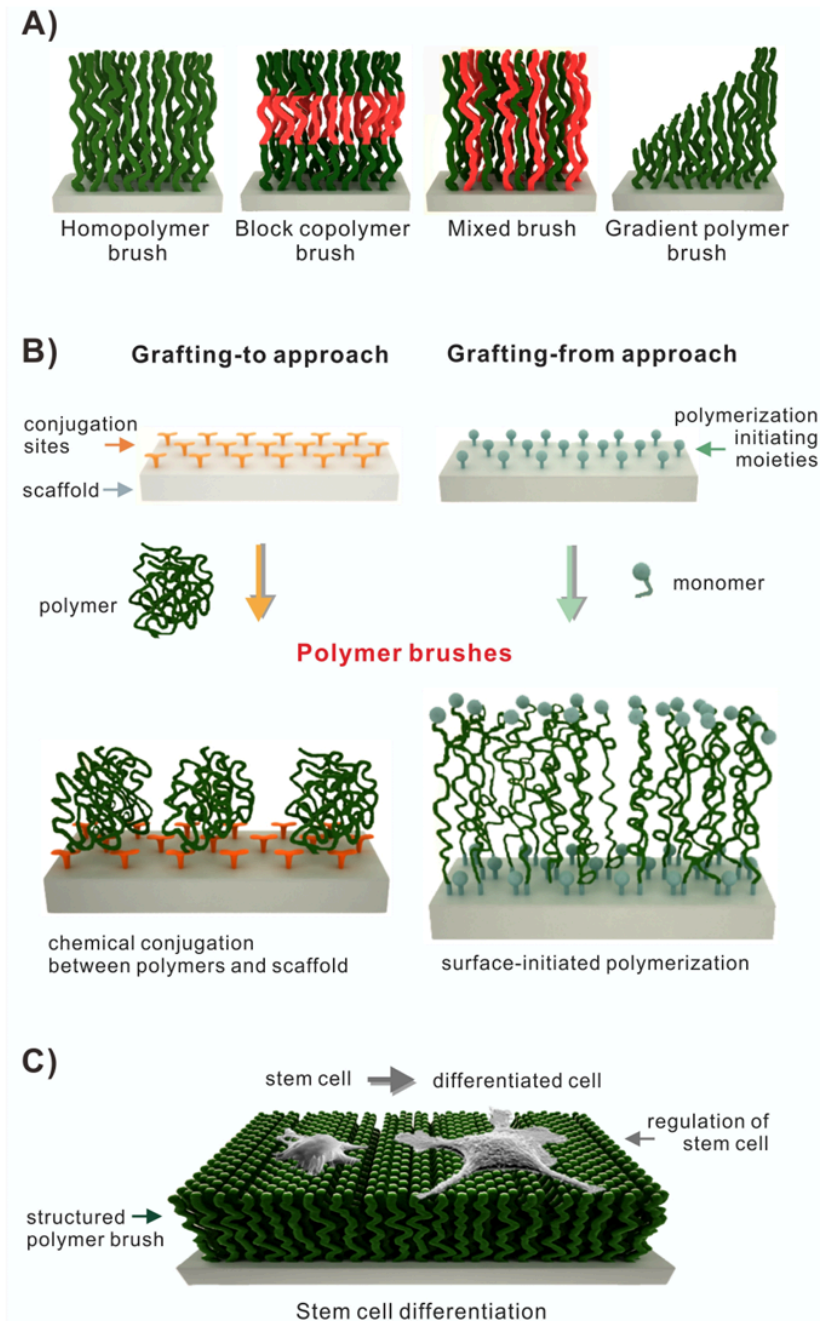

Figure 10. (A) Overview of different types of polymer brushes (homopolymer brush, block copolymer brush, mixed brush, and gradient polymer brush). (B) Fabrication of polymer brushes on scaffolds (Grafting-to approach vs. Grafting-from approach), adapted from ref. 2. (C) Regulation of stem cell differentiation on polymer brush. (Figure and caption from reference [79] with no change under a Creative Commons license.[80])

Table 1. Comparison of the physico-chemical properties between polymer brush and self-assembled monolayer (SAM) (from reference [79] with no change under a creative commons license[80])

| | Polymer brush | | SAM (self-assembled monolayer) |
|---|---|---|---|
| | Grafting-to | Grafting-from | |
| Grafting molecule | Almost all types of polymers | | Mainly alkanethiol & alkyl silnae |
| Micro-architecture | Various and complex polymeric structures | | Well assembled molecular monolayer |
| Scaffold materials | Glass, titanium, gold, silver, silicon, etc | | Gold thin film, oxide-formed substrate |
| Thickness | High tenability by adjusting polymer chain length | | Thin: one molecular layer |
| Coating defects | Presence: short polymer chain Self-healing of defects: long polymer chain | | Presence of defects and pinhole |
| *In vivo* stability | High stability | | Low stability |
| Coating density | Loosely packed | Densely packed | Densely packed |
| Fabrication approach | Various chemical coupling between polymer and surface | Various polymerization on the surface | Thiol-gold bond & silane linkage |

# 5. AI/ML

Machine learning is applied to both simulation and experimentation in scientific discovery.[6,21,81] (Figure 11) Along with reinforcement learning, the major division of ML algorithms is between supervised and unsupervised learning.[8,81,82] In supervised learning, ML can define the relationship between data (inputs) and outputs. Typical statistical methods will be applied, e.g., linear regression, k-nearest neighbors (k-NN), multinomial variation, support vector machines (SVMs), etc., and can be used without training. It relies on good data to be meaningful, which underscores the importance of datasets and literature data from LLMs. In unsupervised learning, ML discovers patterns and relationships that are not apparent, e.g. singular value decomposition (SVD), principal-component analysis (PCA), and autoencoders, which efficiently reduce dimensionality, cluster distribution, and data parametrization. In deep learning (DL), considered a branch or subset of ML, artificial neural networks (ANNs) can be employed for classification, regression, and representation or reinforcement learning (RL).[23,83–86] They are built on layers with interconnected nodes. There are two types of ANN: discriminative and generative models. Generative models enable new data connections. A basic form of discriminative ANN is a multilayer perceptron (MLP).  In convolutional neural networks (CNNs), convolution is used to analyze spatial features with shared-weight filters. They are beneficial for image analysis. Generative adversarial networks (GANs) and variational autoencoders (VAEs) generate new data more efficiently, eventually reaching an equilibrium stage where the discriminator cannot distinguish real from generated data. Deep neural networks (DNNs) have layers between inputs and outputs that can model more complex, non-linear relationships.

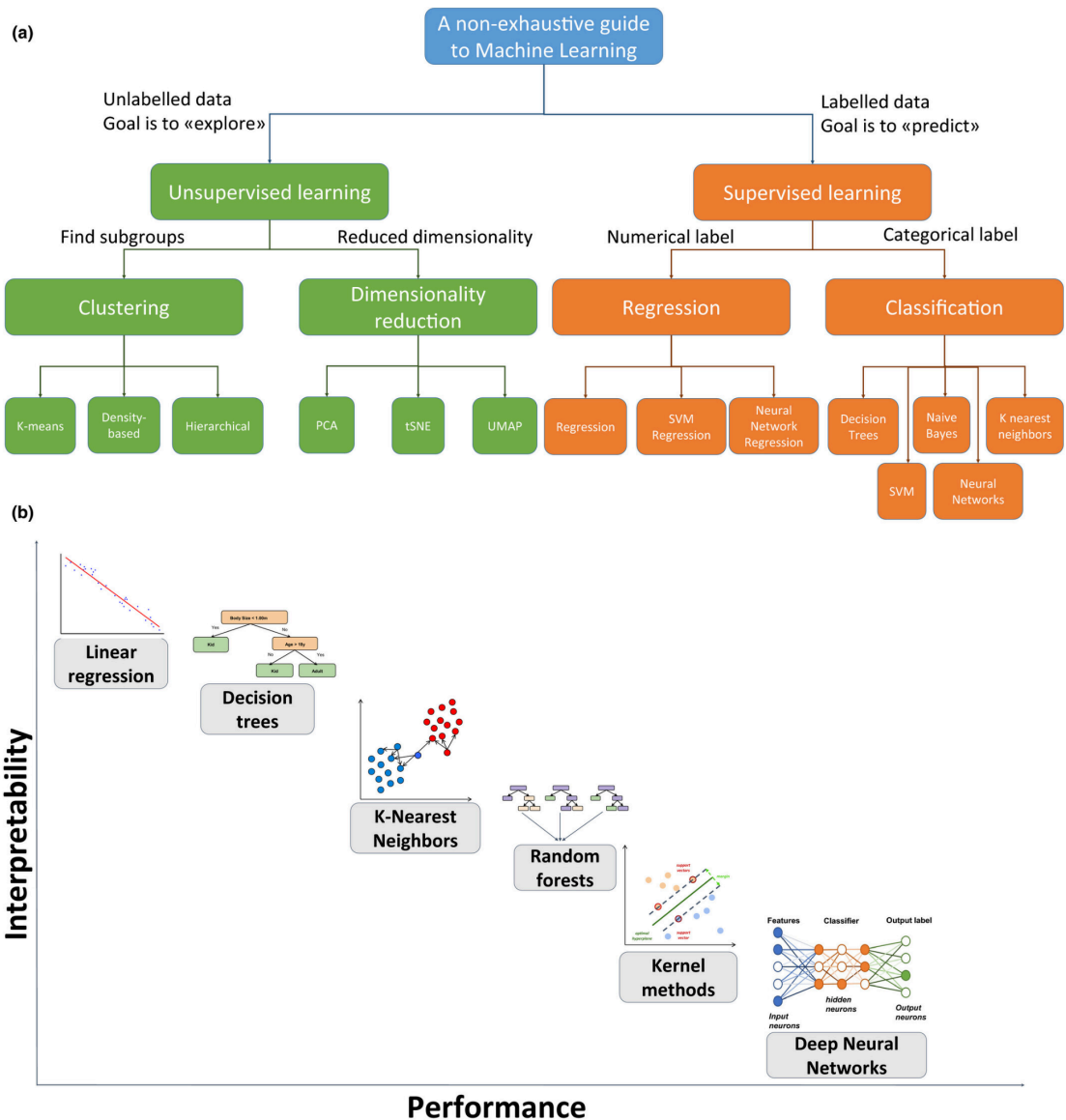

Figure 11. Taxonomy and overview of main machine learning (ML) algorithms. (a) Taxonomy of the different methods presented. (b) Overview of ML methods. The spectrum of available methods ranges from simpler and more interpretable to more advanced algorithms with potentially higher performance at the expense of less interpretability. Position of methods on the figure is qualitative and in practice depends on the number of free parameters, model complexity, data type, and the exact definition of interpretability used. PCA, principal component analysis; SVM, support vector machine; tSNE, t-distributed stochastic neighbor embedding; UMAP, uniform manifold approximation and projection. (Figure and caption from reference,[21] without change, under a creative commons license.[59])

The promising applications of ML in plasmons and optical plasmonics are to accelerate scientific discovery and enhance high performance and adaptability, both in areas of optimization and inverse engineering.[87,88] The objectives are then to introduce ML algorithms and workflows, including supervised, unsupervised, and deep learning, across all aspects of the SPR experiment. The optimal sensitivity and selectivity settings can be identified using large language models (LLMs) and algorithms that can analyze large datasets of sensor responses. Due to Multiphysics and COMSOL's modeling limitations (which can make the process long and complicated, especially with many parameters and variables), ML-optimized workflows can be more efficient.[12,89–91]

There is a large body of reported data that can be fine-tuned using LLMs.[92–94] This will include classification, regression, and dimensionality reduction/clustering based on the existing parametrization. The design of optical parameters and plasmonic fields (evanescent or scattering fields) is governed by established physical principles. ML models can be used as fast surrogate solvers for complex electromagnetic equations (Maxwell's Equations). This will drastically reduce simulation time for predicting optical properties with accelerated protocols towards plug-and-play. The advantage is in rapidly optimizing the S/N and selecting wavelengths that maximize the response. There will be many ML-specific approaches to mining and using LLMs, including plasmon generation, layer architectures, hyphenated microscopy, and nanoparticle-enhanced plasmonic sensors. In ML algorithm development and simulation, statistical models, including Bayesian optimization (BO), linear regression, and decision trees, can be applied to parameter optimization.[8,21,81] The basic optical phenomena for SPR and plasmonics that ML will enhance include absorption, reflection, transmission, emission, and scattering. The emphasis on neural network approaches will accelerate the adoption of DL methods. The use of neural networks and genetic algorithms can enhance the design of more complex plasmonic nanostructures (shape, size, arrangement) and films (layered and complex dielectric designs) to achieve desired optical responses (e.g., specific colors and wavelengths, emission, scattering, etc.). By employing more DL for parameter tuning, digital twins predict optimal sensor geometries and responses to maximize sensitivity and performance in the presence of field effects (electrical, electrochemical, photochemical, pressure, etc.). This can be carried out in the actual device or instrument configuration and planning of the experiment.[1,82]

# 6. Self-Driving Labs

The following are some recent and important examples of ML-driven workflows and applications in SPR instrumentation that augment design, experimental protocols, and predicted data:

Improvement of measuring precision in SPR-based angular scanning is achieved via a DL-based method that can locate the plasmonic angle and boost plasmonic angle detection with no complex post-processing, optical instruments, nor polynomial curve fitting, using a straightforward CNN framework and simulated reflectance spectra as a training dataset (Figure 12).[95]

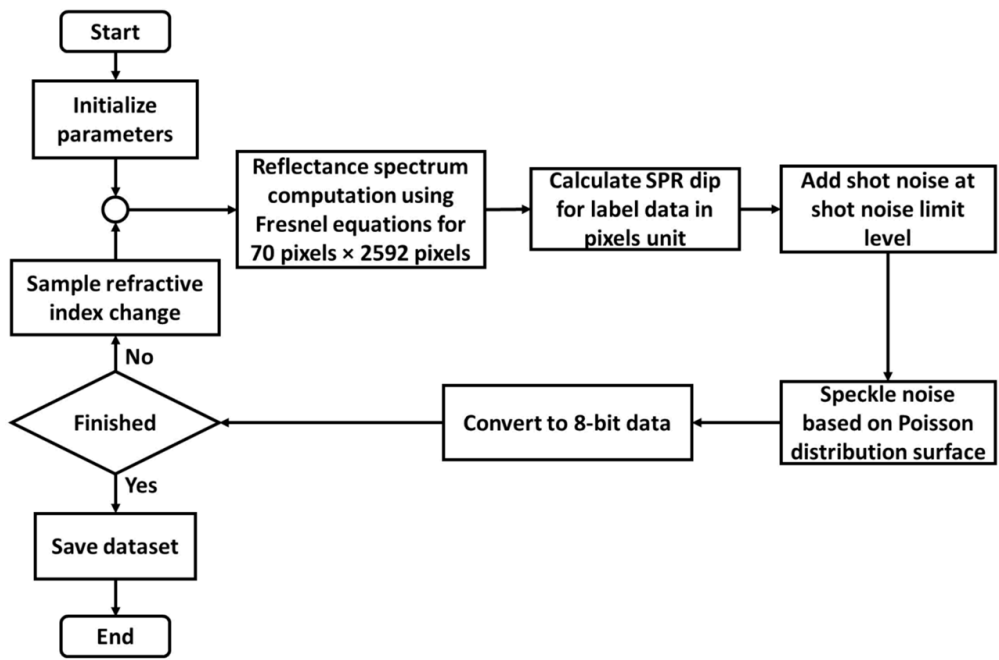

Figure 12. Flowchart of a simulated dataset process. (Figure and caption from reference[95] without change under a Creative Commons license.[9])

An ML-driven CNT-graphene SPR sensor for formalin in water was modeled, built, and experimentally tested, via Linear regression and artificial humming algorithms to determine the optimal geometry variables and SPR sensor layout.[96] A computational tool was utilized for forecasting the SPR response of concave gold nanocubes via a combination of discrete dipole approximation (DDA) and ML methods to calculate the wavelength for dipole SPR of gold concave nanocubes (GCNCs).[97] A regression-based ML model was implemented to pinpoint SPR sensors for enhanced SARS-CoV-2 particle detection both qualitatively and quantitatively.[7] Plasmonic simulations and deep neural networks (DNNs) were deployed to establish a relationship between far-field spectra, near-field distributions, and the dimensional parameters of plasmonic nanospheres, nanorods, and dimers.[98] The SP characteristics of gold sea-urchin-like nanoparticles (GSNPs), promising for cancer thermotherapy, were evaluated with genetic algorithm-driven artificial neural networks (GANNs) to establish the relationship between synthesis variables and surface plasmon wavelengths.[99]

ML was also used to reveal the dynamics of surface refractive index in comb-like plasmonic optical fiber biosensors, with effectiveness across two data-acquisition systems of different resolutions, thus enabling comparative evaluation and increasing the method's robustness.[100]

Besides analyzing, forecasting, or optimizing SPR sensing, AI-enhanced SPR also allows for inverse engineering (Figure 13).[6] Inverse engineering using bidirectional neural networks (a geometry-predicting network and a spectrum-predicting network) can propose a plasmonic structure with the desired transmission spectrum.[6] Furthermore, AI-aided localized surface plasmon resonance (LSPR) was utilized to detect SARS-CoV-2 without sample preparation.[101] The detectable virus concentration was from 125.28 to $10^6$ /mL, with accuracy greater than 97% and $R^2$ > 0.95. (Figure 14)

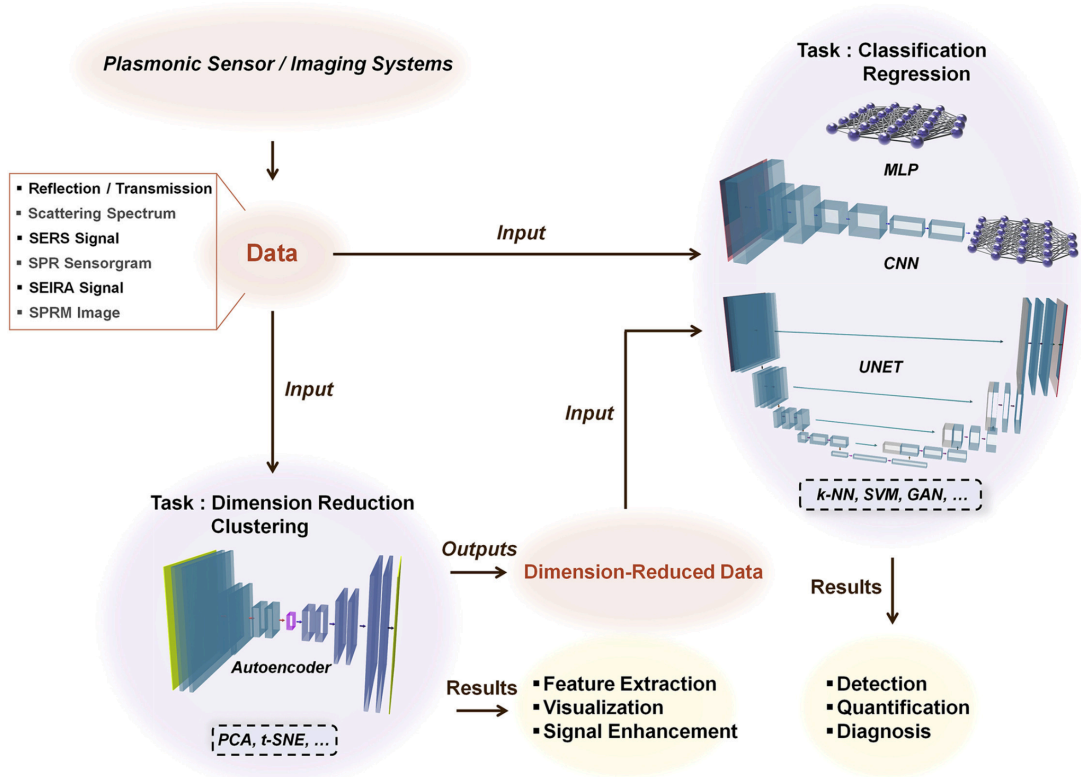

a.

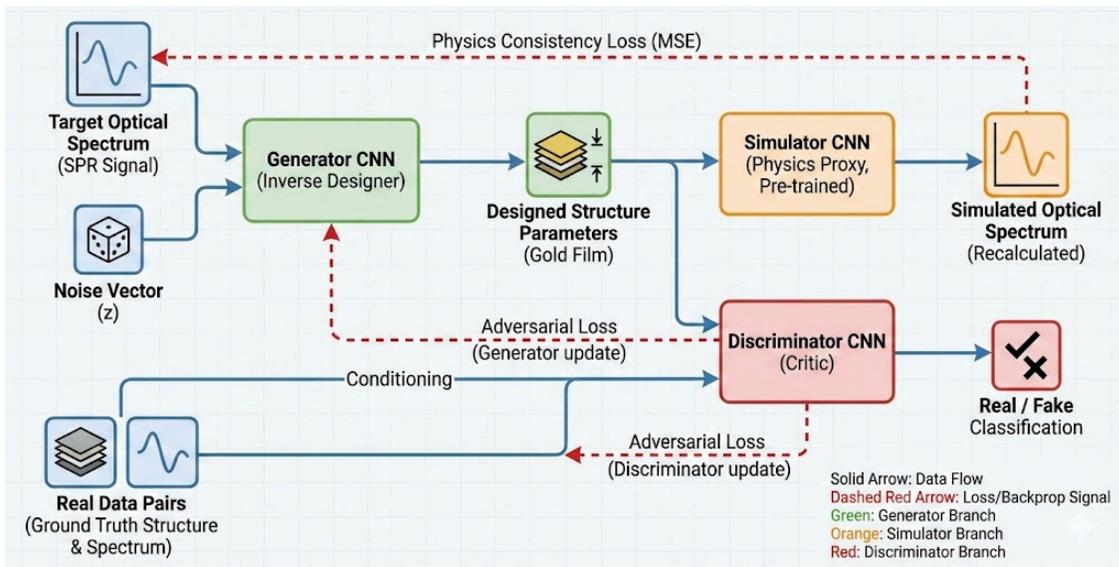

b.

Figure 13. Top: Schematic of ML implementation for analysis of data measured from plasmonic sensing and imaging. (Figure and Caption from reference [6] without change under a Creative Commons license.[9]) Bottom: An architecture of a GAN-based network for designing a gold film structure with targeted optical spectra from inverse engineering of surface plasmon resonance (SPR) signals, using a simulator CNN, generator CNN, and discriminator CNN.

# 7. Future Work and Conclusions

ML-derived workflows in SPR enable improved predictability and solutions based on training data, modeling, and instrument development. With current LLMs trained on existing data and a solid physics foundation, it is possible to generate additional training data, which can then be used to further optimize hardware/software for efficient and consistent data acquisition. This is a virtuous feedback loop mechanism that can be automated with an ML-driven self-driving laboratory (SDL). Nanoparticle and metamaterials plasmonics will be an important direction.[102–106]

**1. Predictability with ML methods**. ML methods will be widely employed to analyze multimodal and high-throughput sensor data. Hyphenated SPR instruments that combine fluorescence, electrochemistry, temperature control, scattering, and microscopy should be made more predictable using ML methods. High-dimensional data obtained can be effectively reduced to a minimum dimensionality, resulting in greater reliability and accuracy for a hyphenated instrument.

**2. ML methods for new dielectric materials and properties.** On the dielectric materials side, simulation and application of ML-optimized molecular dynamics (MD) on the dielectric materials (organic) can be used to understand the more complex interaction between the analyte classes at a finite interface. This can better quantify concentrations and detection limits, even before the experiment. Environmental factors (temperature, humidity, fluid flow, pressure, etc.) can be used to build a multi-parameter-optimized sensor response, which can also be simulated. Lastly, genetic algorithm-based artificial neural networks (GANNs) can be used to guide the synthesis of new dielectric materials, new layer configurations, and new plasmonic nanoparticles with tailored spectral properties for future experiments.

**3. Real-time data and fast computational resources**. ML is less effective when the data is insufficient or when performing relatively simple tasks. The large amount of data can be challenging, and real-time simulation should be possible even with access to powerful high-performance computing (HPC) or exascale computing. Therefore, high-throughput experimentation with an autonomous SDL setup and the generation of well-trained, parametrized data are the most critical paths for development.

**4. Nanoparticles, metamaterials, and photonics.** There are many possibilities for plasmonic structures beyond an ATR-SPR setup. Designing nanostructures and nanoparticles (shapes, size, hot spots, etc.) can be conducted with ANNs and inverse design principles with DL. This is an important frontier that includes metamaterials, auxetic design, and photonics. ML and DL will grow alongside SPR, and plasmonic (including metamaterials) can grow together as powerful tools in materials and analytical chemistry, leading AI/ML in discovery science.

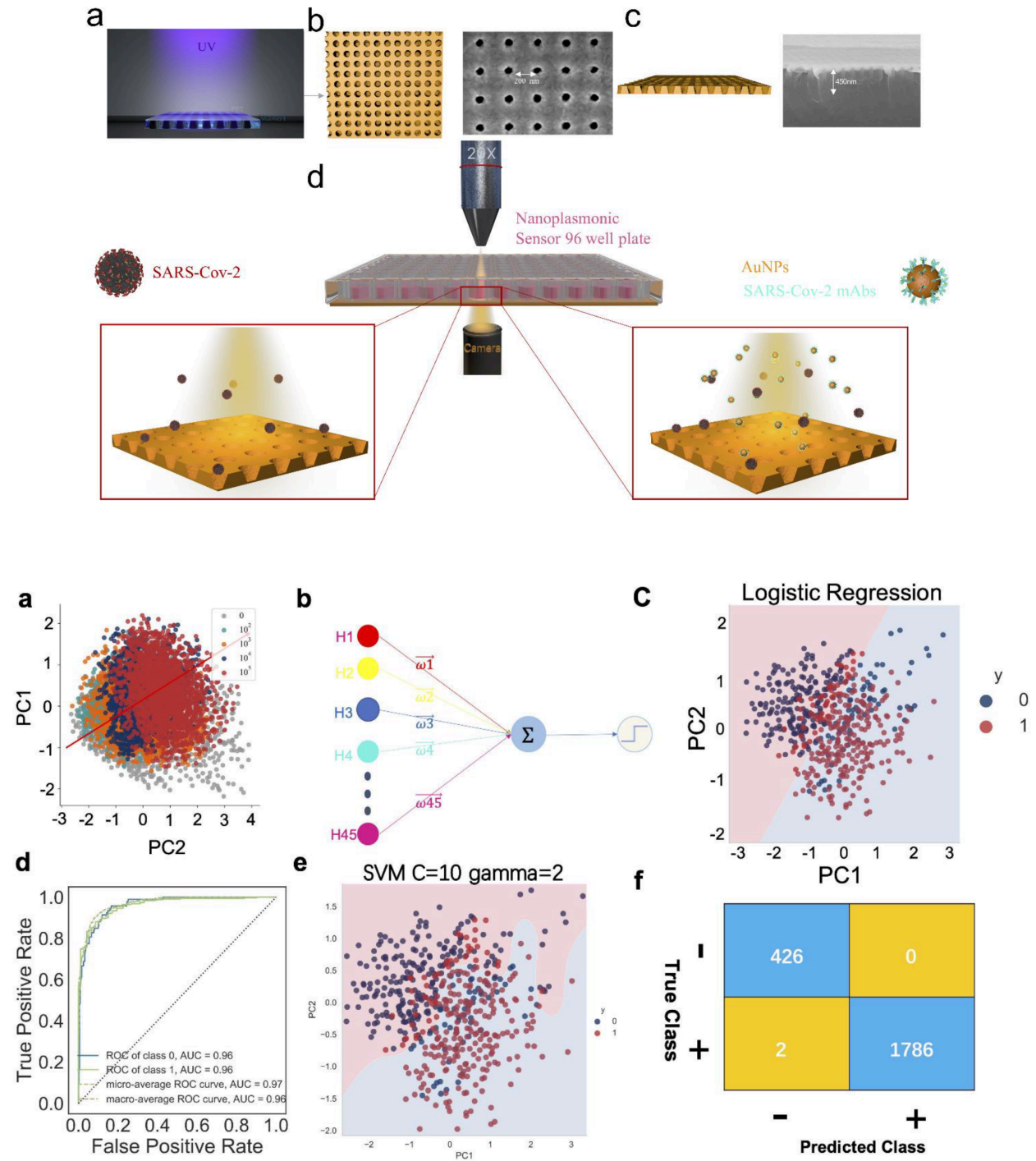

Figure 14. **Top**: Schematic of SARS-CoV-2 detection with plasmonic sensor chip under the microscope. (a) Schematic of the fabrication of Au-TiO2-Au nanocup array chip sensor. Spread UV-curable polymer (NOA61) on the PET sheet, solidify the NOA61 on the PET sheet through UV irradiation, deposit metal on the surface of the NOA61 (left). (b,c) SEM images of the top view

of the chip, SEM images of the cross-sectional view of the chip. (d) Acquire the microscope image of the chip after adding the SARS-CoV-2 virus (left) and acquire the microscope image of the chip after adding the detection antibody (right). **Bottom**: Accurate machine learning classifier models can be obtained from training multiple features. (a) Visualization of the variation in different SARS-CoV-2 concentrations by PCA. (b) The schematic of inputting the ratio values at different degrees of H in different SARS-CoV-2 concentrations into logistic regression model to predict whether it contains the SARS-CoV-2 virus. (c) A schematic diagram of the PC1 and PC2 in logistic regression model. The background color refers to the decision boundary of the classifier (0 refers to the chip image without SARS-CoV-2 virus, and 1 refers to with SARS-CoV-2 virus). (d) The ROC and AUC curve of logistic regression, AUC = 0.95, ROC virus positive of class. (e) A schematic diagram of the PC1 and PC2 in SVM classifier model. The background color refers to the decision boundary of the classifier (0 refers to the chip image SARS-CoV-2 virus negative, and 1 refers to SARS-CoV-2 virus positive, gamma = 2, C = 10). (f) Confusion matrix of SVM classifier (the + refers to virus positive chip image, the − refers to virus negative chip image). (Figures and captions from reference[101] with no change under a creative commons license[9])

## Funding and Acknowledgement

This work was supported by the Center for Nanophase Materials Sciences (CNMS), which is a US Department of Energy, Office of Science User Facility at Oak Ridge National Laboratory, and Laboratory Directed R&D (ORNL INTERSECT).

## Conflicts of Interest

The author declares no conflict of interest.

## Author Contributions

RCA planned the layout. RCA and JC all contributed to the overall design and writing.

**Last updated: 2.16.2026**